\documentclass[fleqn,usenatbib]{mnras}
\usepackage{times}
\usepackage{mathptmx}
\usepackage{subfig}
\usepackage{color}

\usepackage{graphicx}	% Including figure files
\usepackage{amsmath}	% Advanced maths commands
\usepackage{amssymb}	% Extra maths symbols

\newcommand{\degree}{\ensuremath{^\circ}}
\newcommand{\arcsecfrac}{\(\stackrel{\:''}{\textstyle.}\)}

\title[Wolf-Rayet binary in Apep]{Two Wolf-Rayet stars at the heart of colliding-wind binary Apep}
\author[J. R. Callingham et al.]{J.~R.~Callingham,$^{1,2}$\thanks{E-mail: jcal@strw.leidenuniv.nl}
P.~A.~Crowther,$^{3}$
P.~M.~Williams,$^{4}$ 
P.~G.~Tuthill,$^{5}$
Y.~Han,$^{5}$
\newauthor
B.~J.~S.~Pope,$^{6,7,8}$ and
B.~Marcote$^{9}$
\\
$^{1}$Leiden Observatory, Leiden University, PO\,Box 9513, 2300 RA, Leiden, The Netherlands\\
$^{2}$ASTRON, Netherlands Institute for Radio Astronomy, Oude Hoogeveensedijk 4, Dwingeloo, 7991\,PD, The Netherlands\\
$^{3}$Department of Physics \& Astronomy, University of Sheffield, Sheffield, S3 7RH, UK\\
$^{4}$Institute for Astronomy, University of Edinburgh, Royal Observatory, Edinburgh EH9 3HJ, UK\\
$^{5}$Sydney Institute for Astronomy (SIfA), School of Physics, The University of Sydney, NSW 2006, Australia\\
$^{6}$Center for Cosmology and Particle Physics, Department of Physics, New York University, 726 Broadway, New York, NY 10003, USA\\
$^{7}$Center for Data Science, New York University, 60 5th Ave, New York, NY 10011, USA\\
$^{8}$NASA Sagan Fellow\\
$^{9}$Joint Institute for VLBI ERIC, Oude Hoogeveensedijk 4, 7991\,PD Dwingeloo, The Netherlands
}
\date{Accepted 1 May 2020. Received 1 May 2020; in original form 3 April 2020}

\pubyear{2020}

\begin{document}
\label{firstpage}
\pagerange{\pageref{firstpage}--\pageref{lastpage}}
\maketitle

\begin{abstract}

\noindent Infrared imaging of the colliding-wind binary Apep has revealed a spectacular dust plume with complicated internal dynamics that challenges standard colliding-wind binary physics. Such challenges can be potentially resolved if a rapidly-rotating Wolf-Rayet star is located at the heart of the system, implicating Apep as a Galactic progenitor system to long-duration gamma-ray bursts. One of the difficulties in interpreting the dynamics of Apep is that the spectral composition of the stars in the system was unclear. Here we present visual to near-infrared spectra that demonstrate that the central component of Apep is composed of two classical Wolf-Rayet stars of carbon-\,(WC8) and nitrogen-sequence\,(WN4-6b) subtypes. We argue that such an assignment represents the strongest case of a classical WR+WR binary system in the Milky Way. The terminal line-of-sight wind velocities of the WC8 and WN4-6b stars are measured to be $2100 \pm 200$ and $3500 \pm 100$\,km\,s$^{-1}$, respectively. If the mass-loss rate of the two stars are typical for their spectral class, the momentum ratio of the colliding winds is expected to be $\approx$\,0.4. Since the expansion velocity of the dust plume is significantly smaller than either of the measured terminal velocities, we explore the suggestion that one of the Wolf-Rayet winds is anisotropic. We can recover a shock-compressed wind velocity consistent with the observed dust expansion velocity if the WC8 star produces a significantly slow equatorial wind with a velocity of $\approx$530\,km\,s$^{-1}$. Such slow wind speeds can be driven by near-critical rotation of a Wolf-Rayet star.

\end{abstract}


\section{Introduction}
\label{sec:intro}

The luminous, massive Wolf-Rayet (WR) stars are characterised by powerful high-velocity, line-driven winds that carry heavy mass loss \citep{1991ApJ...368..538L}, and are believed to be the immediate precursors to some stripped-envelope core-collapse supernovae \citep{2007ARA&A..45..177C}. A subset of carbon-sequence WR (WC) stars are unique among WR stars since they are often dust-making factories, which can generate spectacular spiral patterns via the interaction with a close companion star \citep{1990MNRAS.243..662W,1999Natur.398..487T}. The complicated dust structures surrounding carbon-rich WR colliding-wind binary (CWB) systems are rare and powerful laboratories for testing our understanding of WR stars as such patterns encode the mass-loss history of the systems. 

The handful of known WR systems with ``pinwheel'' dust patterns, such as archetypal WR\,104 \citep{1999Natur.398..487T} and WR\,98a \citep{1999ApJ...525L..97M}, have demonstrated the morphology of the nebula is linked to the binary orbital motion and mass-loss processes. Since WR stars play a significant role in enriching the interstellar medium \citep{2007ARA&A..45..177C}, and are considered to be the most likely progenitors to long-duration gamma-ray bursts \citep{2006ApJ...637..914W,2008A&A...484..831D,2016SSRv..202...33L}, accurately determining the mass-loss scenario and dynamics of WR stars is integral to improving our understanding of how massive stars die. 

Recently, \citet{2019NatAs...3...82C} discovered an enigmatic source 2XMM\,J160050.7–514245, hereafter called Apep, displaying exceptional X-ray, infrared and radio luminosity for a CWB. The system was identified via infrared imaging to be likely a hierarchical triple, with a massive central CWB that dominates the high-energy, non-thermal, and infrared emission, and a third massive northern companion 0\arcsecfrac7 away. 

The defining feature of Apep is a spectacular 12$''$ mid-infrared dust plume centred on the central binary. While the dust plume strongly resembles the pinwheel nebulae, additional intricate structures suggest that there are unknown processes sculpting the outflow from Apep. In particular, \citet{2019NatAs...3...82C} found that the velocity of the gas in the central binary is a factor of six higher than the velocity derived from the observed proper motion of the dust pattern. As the dust and gas are expected to be coextensive, such a discrepancy between the speed derived from spectra and dust are inconsistent with the existing models of the dynamics of such colliding-wind systems. In standard CWB dust-producing models with isotropic stellar winds, newly formed dust assumes the velocity of the shock compressed stellar wind \citep{Canto1996}, generally $\approx$80 percent of the fastest wind in the system \citep[e.g.][]{Marchenko2003}. The newly formed dust grains are then quickly accelerated to velocities closer to that of the stellar winds in a fraction of an orbital period \citep{pittard2009,williams2009}. All other known Pinwheel nebulae display congruous spectroscopic and dust expansion speeds.

\citet{2019NatAs...3...82C} proposed that the contradiction between the measured dust and gas speeds in Apep can be resolved if the system is capable of launching extremely anisotropic winds, for example due to rapid rotation of the WR star in the central binary. This would imply that Apep is the first known Galactic analogue to long-duration gamma-ray burst progenitors, and a unique laboratory for astrophysics occurring otherwise only at cosmological distances. However, such a conclusion presents its own difficulties as there has not been any definitive observations of critically-rotating WR stars, despite such systems being favoured to exist by long-duration gamma-ray burst models \citep{2006ApJ...637..914W}.

One impediment to our understanding of the dynamics of Apep is the unknown nature of the three stellar components comprising the system. From near-infrared integral-field spectroscopy, the presence of a WR star in the central binary was established by characteristic broad helium and carbon emission lines \citep{2019NatAs...3...82C}, resulting in a Galactic WR catalogue number of WR70-16\footnote{http://pacrowther.staff.shef.ac.uk/WRCat/}. The infrared carbon line ratio diagnostic C\,\textsc{iv}/C\,\textsc{iii} $\lambda$\,1191.1/1199.9\,nm was found to be 3.0, indicating the likely presence of a WC7 type star \citep{Rosslowe2017}. The presence of a dusty late-type WC star was also supported by the observed far-infrared colour excess of Apep.

However, the infrared spectrum of central binary of Apep shows stronger He\,\textsc{ii} and weaker C\,\textsc{iv} line emission than is typical for a WC7 star \citep{Rosslowe2017}. \citet{2019NatAs...3...82C} interpreted the emission line weakness as dilution by additional continuum and, only partly, from warm dust emission. For example, the ratio of the He\,\textsc{ii} $\lambda$\,2189/1163\,nm lines is 0.16 for the central binary of Apep, while other WC7 stars show an average of $\approx$0.38 \citep{Rosslowe2017}. The weakness of the He\,\textsc{ii} emission lines for a WC7 star in the $J$-band, where dust emission is modest, is evidence of additional continuum from a companion star. Based on the abnormal strength of the He\,\textsc{ii} to C\,\textsc{iv} lines for a WC7 star, \citet{2019NatAs...3...82C} suggested that the most likely companion to the WC7 star was an early nitrogen-sequence (WN) WR star. A WN4 or WN5 star was indicated to be the best candidate after comparison to template WN spectra \citep{1996A&A...305..541C}, mainly due to the absence of the N\,\textsc{v} line and relative weakness of the He\,\textsc{i} lines in the near-infrared spectrum.

A binary composed of two classical WR stars located at the heart of Apep would be somewhat surprising since the lifetime of the WR phase is short relative to the evolutionary time of a massive star \citep{2013ApJ...764..166D,2007ARA&A..45..177C}. Based on standard binary evolutionary models, it should be rare to find both stars during such an ephemeral life stage at the same time due to single-star evolution. The (quasi-)\,chemically-homogeneous evolutionary channel is one process that has been suggested as capable of forming a binary composed of two classical WR stars \citep[e.g.][]{martins2013,marchant2016}. In such an evolutionary channel, rapid rotation leads to a mixing timescale shorter than the nuclear timescale, allowing the stars to bypass the post-main-sequence expansion as they would not maintain a massive hydrogen-rich envelope \citep{maeder1987,langer1992}. Smaller radii post-main-sequence implies a binary system could evolve without mass-transfer, providing an opportunity for two massive stars to enter a similar late-evolutionary stage together \citep{song2016}. The rapid rotation required in the model could be sustained via tidal interactions in close binaries \citep{demink2009,mandel2016} but it is unclear how rapid rotation would be maintained in wider binaries.

There are few candidate WR+WR binary systems \citep[e.g.][]{2014MNRAS.445.1663Z,shenar2019}, with the strongest cases being longer-lived main-sequence massive stars exhibiting a WN appearance \citep{2008MNRAS.389L..38S} and HD\,5980 \citep{Koenigsberger2014}. If Apep does harbour a binary comprising two classical WR stars, this will have important consequences for understanding the proposed anisotropic wind model and potential gravitational wave event progenitor systems \citep{2017arXiv170607053B}. 

An alternative spectral class assignment to the WC7+WN4-5 model, which equally-well describes the near-infrared spectra, is that of a WR star in the brief transitory phase between WN and WC phases, labelled WN/C, with an unseen OB-type companion \citep{1989ApJ...337..251C, 1989ApJ...344..870M}. The WN/C classification accurately characterises the line ratios of the C\,\textsc{iii}, C\,\textsc{iv}, and He\,\textsc{i} lines in the near-infrared spectra, and the abnormal strength of the He\,\textsc{ii} lines \citep{Rosslowe2017}. Discerning whether Apep harbours a WN/C is important as the role of this rare and transitory phase between nitrogen-rich and carbon-rich WR types in influencing the total mass-loss of a WR star remains unclear due to the limited number of such systems known. 

One way to distinguish between a WC+WN or a WN/C+O composition is through visual-red spectra, particularly via standard classification
diagnostics, such as C\,\textsc{iii}\,$\lambda$\,569.6\,nm and C\,\textsc{iv}\,$\lambda\lambda$\,580.1-1.2\,nm for WC stars \citep{1998MNRAS.296..367C}. In addition, if the central binary of Apep is a WN/C+O system, all emission lines would be formed in a single wind, whereas one might expect differences in wind velocities for lines arising in each of the WN and WC components \citep{1982MNRAS.198..897W}. Distinction between these two scenarios will allow us to: (1) establish the wind-momentum of the system and wind speeds of the stars in the central binary, (2) calculate the total kinetic power dissipated in the wind collision that is feeding the X-ray and non-thermal radio emission, (3) probe the geometry of the wind-collision region, which in turn shapes the larger dust plume, and; (4) derive an accurate distance to the system based on spectral class luminosities. Additionally, confident identification of the components of the central binary of Apep will facilitate planning and interpretation of high resolution optical aperture-masking and very long baseline interferometry (VLBI) radio observations. 

With these goals in mind, we observed Apep with the X-SHOOTER spectrograph \citep{2011A&A...536A.105V} on the European Southern Observatory’s (ESO's) Very Large Telescope (VLT). The X-SHOOTER observations and the data reduction process performed are provided in Section\,\ref{sec:obs}. Section\,\ref{sec:results} details the spectral line and type classification of the central binary and third massive member of the Apep system. The implications of the X-SHOOTER spectra on our understanding of the Apep system is discussed in Section\,\ref{sec:discuss}, and our study is summarised in Section\,\ref{sec:concl}.  

\section{X-SHOOTER Observations and Data Reduction}
\label{sec:obs}

We observed Apep using the X-SHOOTER spectrograph \citep{2011A&A...536A.105V} on UT2 at ESO's VLT for a total of 45\,minutes on 2019 April 9 (Program ID: 0103.D-0695(A); PI: Callingham). While X-SHOOTER is simultaneously sensitive from ultraviolet-blue to near-infrared (300 to 2500\,nm), the setup of the observation was optimised to extract information from the visual-red arm (550 to 1000\,nm) of the spectrograph. Therefore, no nodding was performed. Such a choice was made because reliable near-infrared spectra already exist for the system, and the dust extinction towards Apep makes the ultraviolet-blue data so faint as to be unusable without significantly more observing time.

The visual-red and near-infrared arm were set to have a slit width of 0.9$''$, resulting in a nominal spectral resolution of 8800 and 5100, respectively. The slit was positioned along the axis connecting the central binary and the third massive northern companion. Exposures were of $\approx$300 seconds duration. The observations were conducted in seeing conditions of $\approx$0\arcsecfrac7 and with an airmass of $\approx$1.2. We also observed the standard star $\theta$\,Normae (HIP\,79653) to correct for telluric spectral features.

To perform the data reduction we followed the standard X-SHOOTER pipeline \citep[v3.2.0;][]{2010SPIE.7737E..28M} via \textsc{EsoReflex}. Bad pixel maps, bias correction, flat-fielding, detector linearity, and wavelength calibration from standard-lamps were applied to $\theta$\,Normae and Apep frames for each arm of the spectrograph. Each scan was combined, and the median taken, to form the highest signal-to-noise spectra. There was no usable science data in the ultraviolet-blue arm, so it will not be discussed further.

To isolate the spectrum of the inner binary from the northern companion, the size of the extraction aperture was adjusted to isolate the broad component of the 656\,nm line for the visual-red spectrum and the He\,\textsc{i}\,$\lambda$\,1083\,nm line for the near-infrared spectrum. The 656\,nm line corresponds to He\,\textsc{ii}+C\,\textsc{iv} and H\,$\alpha$ for the central binary and northern companion, respectively. The broad component of these lines are produced by the WR star(s). The reverse was performed to isolate the spectrum of the northern companion. Based on the fine structures present in the central binary's He\,\textsc{i}\,$\lambda$\,1083\,nm line, we find cross-contamination at long wavelengths to be $< 5\%$, and smaller at shorter wavelengths.

$\theta$\,Normae's spectra was also extracted and then normalised by a black body curve of the appropriate temperature. The standard star's intrinsic spectral features were removed via modelling of the lines with Lorentzian profiles. The resulting standard star spectra were then used to correct for telluric features in the science spectra of the central binary and northern companion. These final science spectra in each waveband were then continuum-corrected.

\section{Spectroscopic results}
\label{sec:results}

\subsection{Composition of the central binary of Apep}\label{central}

\begin{figure*}
\begin{center}
\includegraphics[scale=0.6]{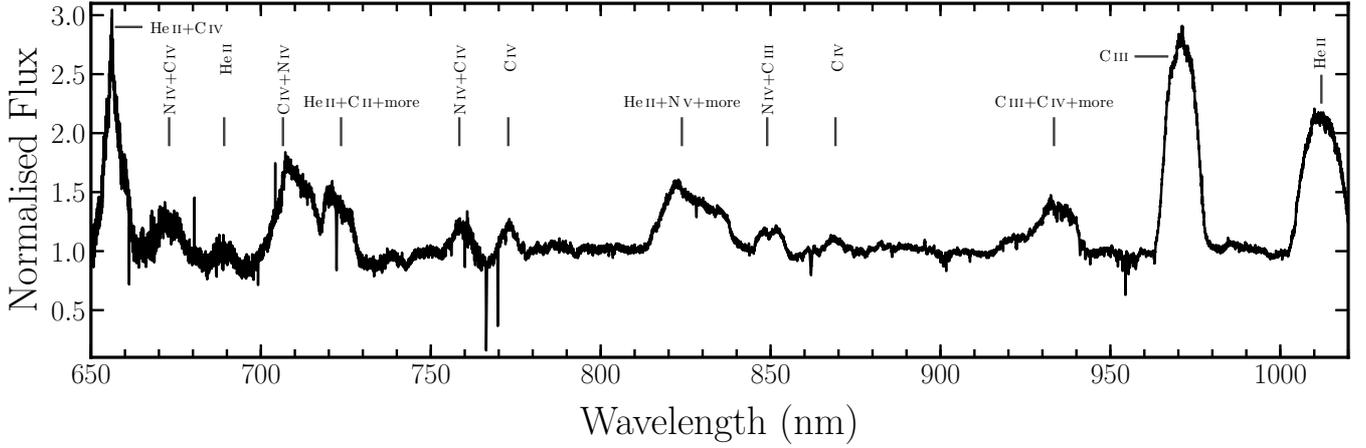}
 \caption{The spectrum of the central binary of Apep from 650 to 1020\,nm. Prominent emission lines are labelled, with all detected emission lines provided in Table\,\ref{tab:line_id}.} 
\label{fig:cent_binary_vis}
\end{center}
\end{figure*}

\begin{figure}
\begin{center}
\includegraphics[scale=0.45]{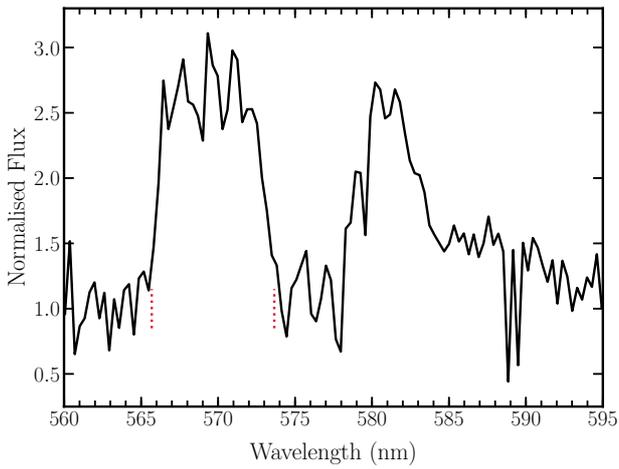}
 \caption{The spectrum of the central binary of Apep from 560 to 590\,nm. The C\,\textsc{iii}\,$\lambda$\,569.6\,nm and C\,\textsc{iv}\,$\lambda\lambda$\,580.1-1.2\,nm emission lines are detected in this wavelength range. The asymmetric structure of the C\,\textsc{iv}\,$\lambda\lambda$\,580.1-1.2\,nm doublet is likely due to the presence of the He\,\textsc{i}\,$\lambda$\,587.6\,nm line. The spectral resolution of this plot has been smoothed by a factor of 16 to increase the signal-to-noise of the two lines. The strong Na\,\textsc{i} D interstellar doublet around 589\,nm is also detected, as are diffuse interstellar bands at 578 and 579.9\,nm. The dotted-red lines correspond to the full-width at zero-intensity (FWZI) of the C\,\textsc{iii}\,$\lambda$\,569.6\,nm line.} 
\label{fig:vis_clines}
\end{center} 
\end{figure}

\begin{figure}
\begin{center}
\includegraphics[scale=0.45]{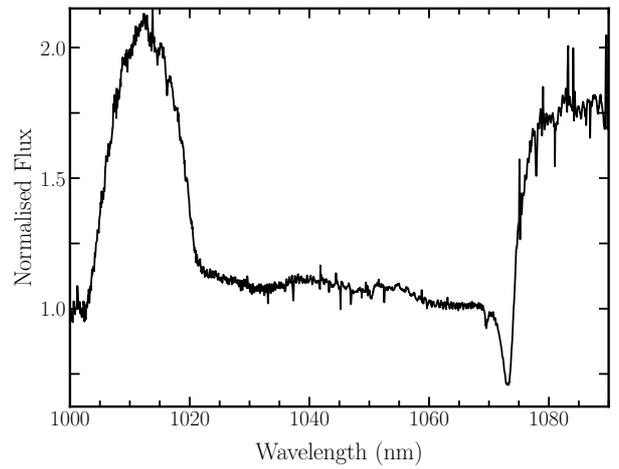}
 \caption{The spectrum of the central binary of Apep from 1000 to 1090\,nm. The two lines detected are He\,\textsc{ii}\,$\lambda$\,1012\,nm, present also in Figure\,\ref{fig:cent_binary_vis}, and He\,\textsc{i}\,$\lambda$\,1083\,nm. We do not present spectra further than 1090\,nm as sky lines begin to be become more prominent and bad columns prevent data being recovered between 1090 and 1096\,nm.} 
\label{fig:cent_binary_nir}
\end{center}
\end{figure}

We present the X-SHOOTER spectrum of the central binary of Apep from 650 to 1020\,nm in Figure\,\ref{fig:cent_binary_vis}. The data shortward of 650\,nm was of low signal-to-noise, and therefore was binned by a factor of sixteen to aid in the detection of broad lines. The two detected lines in that wavelength range, namely the primary WC classification diagnostics C\,\textsc{iii}\,$\lambda$\,569.6\,nm and C\,\textsc{iv}\,$\lambda\lambda$\,580.1-1.2\,nm, are shown in Figure\,\ref{fig:vis_clines}. We present the usable data between 1000 and 1090\,nm from the near-infrared arm in Figure\,\ref{fig:cent_binary_nir}. Data longward of 1090\,nm had issues with bad columns and prominent sky lines, and was already presented by \citet{2019NatAs...3...82C}. All detected emission lines and their equivalent widths (EWs) are presented in Table\,\ref{tab:line_id}.

\begin{table}
\caption{\label{tab:line_id} Observed wavelength ($\lambda_{\text{obs}}$), equivalent width (EW), and identification of the detected emission lines \citep{2018Galax...6...63V} in the X-SHOOTER visual to near-infrared spectra of the central binary of Apep. Blended lines for which we can not split are identified by `(bl)', with the line that likely contributes the most to the reported EW listed first. The Carbon line contributors to the 826\,nm blend are 826.3 C\,\textsc{iii} 8$h$-6$g$; 828.6 C\,\textsc{iii} 6$d$-5$f$; 831.4 
C\,\textsc{iii} 8$i$-6$h$; 831.9 C\,\textsc{iii} 6$d$-3$d$; and 832.3 C\,\textsc{iii} 6$p$-3$p$.}
\begin{center}
\begin{tabular}{lll}
\hline
$\lambda_{\text{obs}}$ &  EW  & Identification                             \\
 (nm)  &  (nm)  & (nm) ions, multiplets, transition arrays   \\
\hline
 569.5 & 10.38\,$\pm$\,0.07  & 596.7 C\,\textsc{iii} 3$d$-3$p$                         \\
 581.2 &  5.63\,$\pm$\,0.09  & 580.3 581.5 C\,\textsc{iv} 3$p$-3$s$                    \\
 587.7 &  0.5\,(bl)  & 587.7 He\,\textsc{i} 3$d$-2$p$                          \\
 656.4 &  8.88\,$\pm$\,0.02  & 656.2 He\,\textsc{ii} (6-4), C\,\textsc{iv} (12-8)            \\
 673.0 &  1.82\,$\pm$\,0.07  & 674.0  N\,\textsc{v} 8p-7s;  674.8 C\,\textsc{iv} (16-9)\\ 
 689.2 &  0.9\,$\pm$\,0.1  & 689.3 He\,\textsc{ii} (12-5)                                    \\
 709.1 & 11.2\,(bl)  & 706.5 C\,\textsc{iv} (9-7); 711.1-712.7 N\,\textsc{iv} 3$d$-3$p$\\
 722.8 &  4.4\,(bl) & 718.0 He\,\textsc{ii} (11-5); 720.6 N\,\textsc{iv} 4$d$-4$p$; \\
 & &  721.3 C\,\textsc{iii} 6$d$-5$p$; 723.8 C\,\textsc{ii} 3$d$-3$p$  \\
 739.0 &  0.4\,$\pm$\,0.1  & 738.4 C\,\textsc{iv} 7$d$-6$p$ \\
 759.2 &  1.77\,$\pm$\,0.05  & 758.4 N\,\textsc{iv} 7$h$-6$g$; 758.7 C\,\textsc{iv} 11$d$-8$p$; \\
 & & 759.4 C\,\textsc{iv} (20-10)                     \\
 773.1 &  1.36\,$\pm$\,0.05  & 772.8 C\,\textsc{iv} (7-6)                                   \\
 826.1 &  9.07\,$\pm$\,0.06  & 823.9 He\,\textsc{ii} (9-5); 822.4 N\,\textsc{v} 16$d$-11$p$;\\
 & & many\,C\,\textsc{iii} (listed above)\\         
 847.8 &  0.6\,(bl) & 848.4 N\,\textsc{iv} 10$p$-8$s$                                  \\
 851.9 &  0.9\,(bl) & 850.3 C\,\textsc{iii} 3$p$-3$s$; 852.0 He\,\textsc{i} 8$p$-3$s$               \\
 868.6 &  0.3\,$\pm$\,0.1  & 869.1 C\,\textsc{iv} (17-10)                                 \\
 934.5 &  3.41\,$\pm$\,0.07  & 933.4 C\,\textsc{iii} 2$p$.3$p$-2$s$.4$p$; 933.8 C\,\textsc{iv} (16-10); \\
 & & 937.0 He\,\textsc{ii} (17-6)  \\
 970.9 & 17.43\,$\pm$\,0.03  & 970.8 C\,\textsc{iii} 3$d$-3$p$; 971.8 C\,\textsc{iii} 3$d$-3$p$                \\
1012.1 &  22.65\,$\pm$\,0.04 & 1012.6 He\,\textsc{ii} (5-4); 1012.4 C\,\textsc{iv} (10-8)               \\
1054.5 &  0.4\,$\pm$\,0.1  & 1054.5 C\,\textsc{iv} (12-9)                                   \\
1086.0 &  NA      & 1083.3 He\,\textsc{i} 2$p$-2$s$               \\
% 1163.2 & 14.25  & 1163.0  He\,\textsc{ii} (7-5), 1162.9 C\,\textsc{iv} (14-10), 1164.6 C\,\textsc{iii} 7$d$-6$p$ \\
% 1191.2 &  5.8:  & 1190.9  C\,\textsc{iv} (8-7)                                \\
% 1201.5 &  1.5:  & 1199.1  C\,\textsc{iii} 4$p$-4$s$                              \\
% 1253.3 &  1.33  & 1253.1  C\,\textsc{iii} (9-7), 1254.5 C\,\textsc{iii} (7-6)          \\
% 1280.4 &  5.11  & 1281.6  He\,\textsc{ii} (10-6), 1278.4 He\,\textsc{i} (5-3)            \\
\hline
\end{tabular}
 \end{center}
\end{table}

To determine whether the central binary is composed of WC+WN stars or a WN/C with an unseen OB-companion, we consider the equivalent width ratio of the C\,\textsc{iii}\,$\lambda$\,971\,nm and He\,\textsc{ii}\,$\lambda$\,1012\,nm lines relative to those of C\,\textsc{iv}\,$\lambda$\,1191\,nm/He\,\textsc{ii}\,$\lambda$\,1163\,nm and C\,\textsc{iv}\,$\lambda$\,2078\,nm/C\,\textsc{iii}+He\,\textsc{i}\,$\lambda$\,2117\,nm \citep{2019NatAs...3...82C}. We find that the C\,\textsc{iii}\,$\lambda$\,971\,nm and He\,\textsc{ii}\,$\lambda$\,1012\,nm line ratio is $\approx$0.44. Such a line ratio suggests that the central binary is similar to the WN/C stars WR\,8, WR\,26, and WR\,98 \citep{Rosslowe2017}.

Aside from the strong C\,\textsc{iii}\,$\lambda$\,971\,nm and He\,\textsc{ii}\,$\lambda$\,1012\,nm lines, prominent features arising in WN and WN/C stars include the N\,\textsc{iv}\,$\lambda$\,711.6\,nm \citep{1968MNRAS.138..109S}, which is also present in the spectrum of the central binary of Apep. In comparison, prominent features in WC and WN/C stars are the C\,\textsc{iii}\,$\lambda$\,674\,nm, C\,\textsc{iv}\,$\lambda$\,706.5\,nm, and C\,\textsc{iii}\,$\lambda$\,850\,nm lines \citep{2007ARA&A..45..177C}. As shown in Figure\,\ref{fig:cent_binary_vis}, the C\,\textsc{iii}\,$\lambda$\,674\,nm line is present, while the C\,\textsc{iv}\,$\lambda$\,706.5\,nm line is blended with N\,\textsc{iv}\,$\lambda$\,711.6\,nm, and the C\,\textsc{iii}\,$\lambda$\,850\,nm is relatively weak. Several line profiles are extremely broad, suggesting an extremely high wind velocity for at least one WR star in Apep \citep{1994MNRAS.269.1082E}.

\begin{figure}
\begin{center}
\includegraphics[scale=0.35]{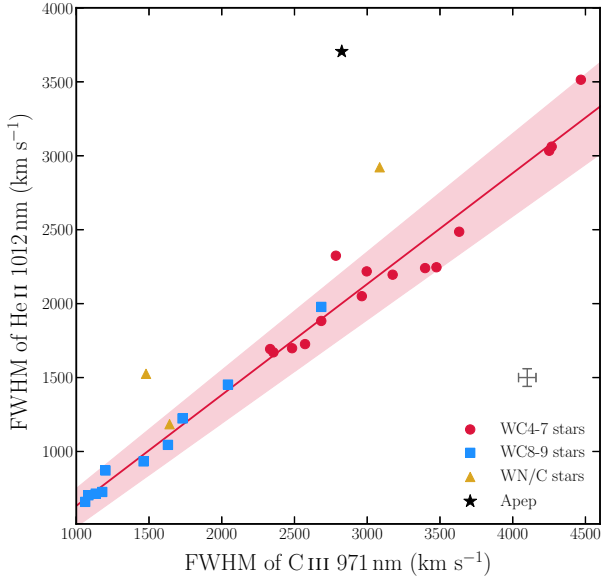}
 \caption{The FWHM of the He\,\textsc{ii}\,$\lambda$\,1012\,nm and C\,\textsc{iii}\,$\lambda$\,971\,nm lines for single WC4-7, WC8-9, and WN/C stars \citep{Rosslowe2017}, as represented by red circles, blue squares, and yellow triangles, respectively. The central binary of Apep is shown as a black star, and the uncertainties derived from Gaussian fits to the respective emission lines are approximately the size of the symbol. The red line is the best fit to the literature data for the WC stars: FWHM(He\,\textsc{ii} 1012) $ = (0.75 \pm 0.05)$FWHM(C\,\textsc{iii} 971)$ - (120\pm70)$\,km\,s$^{-1}$. The shaded region represents the 3-$\sigma$ uncertainty around this fit. The median uncertainty of the literature FWHM measurements is shown by the gray errorbars above the legend. The two WN/C stars offset from the trend are likely outliers due to abnormal chemical abundances, and do not exceed a 1:1 FWHM ratio for the two lines.} 
\label{fig:trend_ciii_heii}
\end{center}
\end{figure}

\begin{figure}
\begin{center}
\includegraphics[scale=0.35]{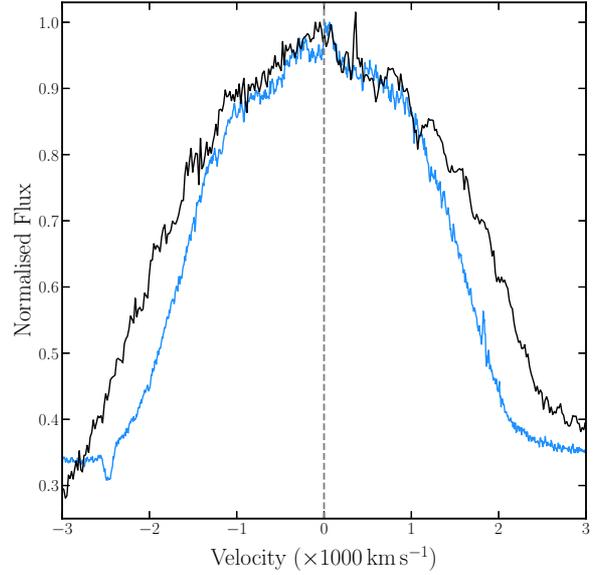}
 \caption{Comparison of the profiles of the He\,\textsc{ii}\,$\lambda$\,1012\,nm and C\,\textsc{iii}\,$\lambda$\,971\,nm lines, shown in black and blue, respectively. To aid in a visual comparison of the FWHM of the two lines, the height of the lines are normalised by their respective peak flux near zero velocity. The FWHM for the He\,\textsc{ii}\,$\lambda$\,1012\,nm line is larger than the C\,\textsc{iii}\,$\lambda$\,971\,nm line, as expected if the lines are formed in two different stars. The data of the He\,\textsc{ii}\,$\lambda$\,1012\,nm and C\,\textsc{iii}\,$\lambda$\,971\,nm lines are from the near-infrared and visual-red arms of the X-SHOOTER spectrograph, respectively. } 
\label{fig:fwhm_ciii_heii}
\end{center}
\end{figure}

In order to break the apparent degeneracy in spectral morphology between WN/C or WN+WC components, a clear expectation of the single WN/C star scenario is that the emission lines from ions of high ionisation energy form in the inner, accelerating wind, while lower ionisation lies form in the outer wind moving close to terminal velocity \citep{1982MNRAS.198..897W, 1989ApJ...347..392H}. Therefore, the velocities of the different ions need to be physically consistent if the wind lines arise from a single star. For example, He\,\textsc{ii} lines form at a smaller radii than C\,\textsc{iii} lines, implying we should expect the full-width half-maximum (FWHM) of He\,\textsc{ii} lines to be smaller than C\,\textsc{iii} lines. Indeed, we find from a large sample of single WC4-9 stars the trend: FWHM(He\,\textsc{ii} 1012)/FWHM(C\,\textsc{iii} 971) = 0.75$\pm$0.05 \citep{Rosslowe2017}, as shown in Figure\,\ref{fig:trend_ciii_heii}. In contrast, Apep is a significant outlier in this population with a FWHM(He\,\textsc{ii} 1012)/FWHM(C\,\textsc{iii} 971) = 1.30\,$\pm$\,0.05. We also directly demonstrate in  Figure\,\ref{fig:fwhm_ciii_heii} that the FWHM of the  He\,\textsc{ii}\,$\lambda$\,1012\,nm line exceeds that of the C\,\textsc{iii}\,$\lambda$\,971\,nm line in Doppler space. We interpret such an unusual ratio as implying that these lines do not arise in a single WN/C star. Therefore, we suggest that the central binary of Apep is composed of a double WR system.

\begin{figure*}
\begin{center}
\includegraphics[scale=0.55]{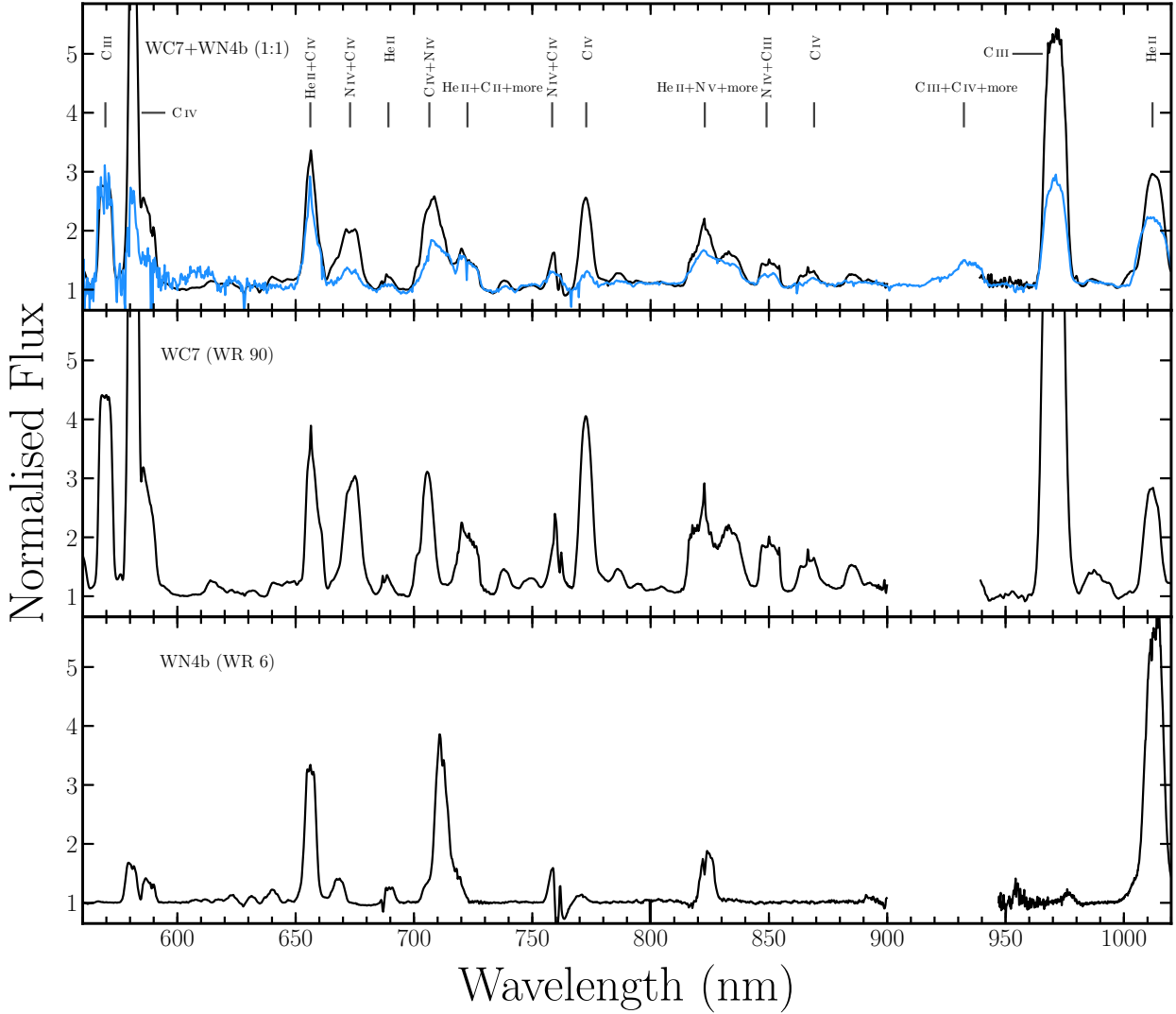}
 \caption{Visual-red spectra of a WN4b star \citep[WR\,6; bottom panel;][]{hamann1995,Howarth1992}, a WC7 star \citep[WR\,90; middle panel;][]{dessart2000,Rosslowe2017}, and their identical combination (1:1; top panel). The visual-red X-SHOOTER spectrum of Apep's central binary (also presented in Figures\,\ref{fig:cent_binary_vis} and \ref{fig:vis_clines}) is overplotted in the top-panel in blue, with the spectral resolution smoothed by a factor of 16. The close morphological agreement of the composite spectrum to the central binary of Apep qualitatively supports the WC8+WN4-6b composition, and that host dust is diluting the near-infrared lines. Prominent emission lines are labelled similarly to Figure\,\ref{fig:cent_binary_vis}.}
\label{fig:comp}
\end{center}
\end{figure*}

Such an assignment is further supported by the spectra presented in Figure\,\ref{fig:comp}, which demonstrates that the identical combination of the spectra of a WC7 \citep[WR\,90;][]{dessart2000,Rosslowe2017} and a WN4b \citep[WR\,6;][]{hamann1995,Howarth1992} star displays a close qualitative agreement to the visual-red spectra of the central binary of Apep, as provided in Figure\,\ref{fig:cent_binary_vis} and \ref{fig:vis_clines}. The morphological agreement of the spectra produced by assuming an identical contribution  of WR\,90 and WR\,6 to the continuum also indicates that neither WR star in central binary of Apep strongly dominates the visual-red spectra, and hot dust is diluting the near-infrared emission lines.

The conclusion of a WC+WN system assumes that the reported line ratios are clean proxies for the ion stratification in the stellar winds. We do note that there are several known mechanisms that could distort the line ratios presented. For example, the profile of the C\,\textsc{iii} lines are susceptible to the geometry and dynamics of the wind collision region in a CWB \citep{luehrs1997}. However, an excess in the C\,\textsc{iii}  lines for CWBs has only been observed in systems with short (days to months) periods \citep[e.g.][]{hill2000,bartzakos2001}. In comparison, the central binary of Apep has an orbital period of $\approx$100\,yr \citep{2019NatAs...3...82C} and is not actively producing dust \citep{han2020}. The long period of Apep also implies the wind-collision region is unlikely to be disrupting the ionisation stratification of the stellar winds. Furthermore, the relatively low inclination of the Apep system argues against a significant amount of compressed wind velocity in our line of sight. If the wind-collision region was contributing significantly to the C\,\textsc{iii} lines, we would expect to observe prominent sub-peak structure to be evident in the C\,\textsc{iii}\,$\lambda$\,569.6\,nm line. Such sub-peak structure would likely be at relatively low radial-velocity given the geometry of the system, and only strong when dust formation is vigorous. 

Since we conclude that the central binary is likely composed of two classical WR stars, we can use the carbon and nitrogen line diagnostics to pin down their spectral subtypes. The C\,\textsc{iii}\,$\lambda$\,569.6\,nm and C\,\textsc{iv}\,$\lambda\lambda$\,580.1-1.2\,nm lines are weak or absent in early WN stars. Therefore, we can classify the WC component from their relative strengths. As shown in Figure\,\ref{fig:vis_clines}, both lines are detected and have an equivalent width ratio of C\,\textsc{iii}\,$\lambda$\,569.6\,nm/C\,\textsc{iv}\,$\lambda\lambda$\,580.1-1.2\,nm $\approx$\,1.8, establishing the WC component as a WC8 star \citep{1998MNRAS.296..367C}. The C\,\textsc{iii}\,$\lambda$\,1191\,nm/C\,\textsc{iv}\,$\lambda$\,1199\,nm ratio of $\approx$3.0 \citep{2019NatAs...3...82C} argues against a WC9 classification. 

Equivalent widths for the WN star, using the strength of the helium and nitrogen lines, implies the WN component is consistent with a broad-lined WN4-6 star \citep{1996MNRAS.281..163S, Rosslowe2017}, with the broad-line assignment supported by FWHM(He\,\textsc{ii}\,1012) $\gg$ 1900 km\,s$^{-1}$. In particular, the detection of the N\,\textsc{iv}\,$\lambda$\,712\,nm line narrows the spectral type of the WN star to WN4-6 as the N\,\textsc{iv}\,$\lambda$\,712\,nm line is absent or weak in both WN2-3 and WN7-9 stars \citep{1996MNRAS.281..163S, Rosslowe2017}. Hence, we suggest that the central binary of Apep is composed of a WC8+WN4-6b binary.

For a WC8+WN4-6 binary, we would expect each component to contribute more or less equally to the continuum flux in the visible range \citep{2020MNRAS.493.1512R}, as also supported by the close morphological agreement of the combined WR\,90+WR\,6 spectrum in with that of the central binary (Figure\,\ref{fig:comp}). Since the WC8 component is expected to dominate both the C\,\textsc{iii}\,$\lambda$\,569.6\,nm and C\,\textsc{iv}\,$\lambda\lambda$\,580.1-1.2\,nm lines, the intrinsic equivalent widths of these lines in the WC8 star are expected to be a factor of two larger than observed. Such strengths are relatively modest with respect to other WC8 stars. For example, WR\,60 (HD~121194) has a similar FWHM to the WC8 component of Apep but has the equivalent widths of the C\,\textsc{iii}\,$\lambda$\,569.6\,nm and C\,\textsc{iv}\,$\lambda\lambda$\,580.1-1.2\,nm lines a factor of three times larger \citep{1990ApJ...358..229S}. Therefore, it is likely the WC8 star has unusually weak lines to allow the WN4-6b star to contribute significantly more than expected to the visible continuum flux. We note that the weak line abnormality for the WC8 star can also suggest an additional continuum source, as hot dust will not be prominent at visible wavelengths. However, the internal dynamics and smooth pinwheel dust pattern of Apep does not support a close third member in the central component. 

\subsection{Classification of the northern companion}

\begin{figure*}
\begin{center}
\includegraphics[scale=0.6]{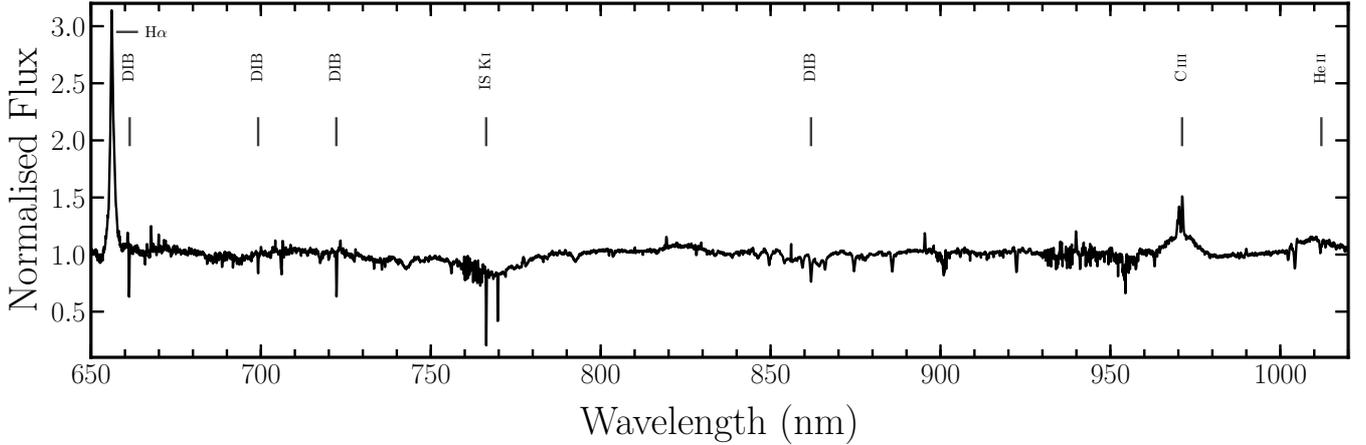}
 \caption{The spectrum of the northern companion to the central binary of Apep from 650 to 1020\,nm. Emission lines that are prominent are labelled, plus members of the Paschen series are seen (P$\delta$, P$\epsilon$ etc.). Known diffuse interstellar absorption bands and interstellar potassium lines are labelled by `DIB' and `IS\,K\textsc{i}', respectively, and are also present in the spectrum of the central binary shown in Figure\,\ref{fig:cent_binary_vis}.} 
\label{fig:north_com_vis}
\end{center}
\end{figure*}

\begin{figure}
\begin{center}
\includegraphics[scale=0.45]{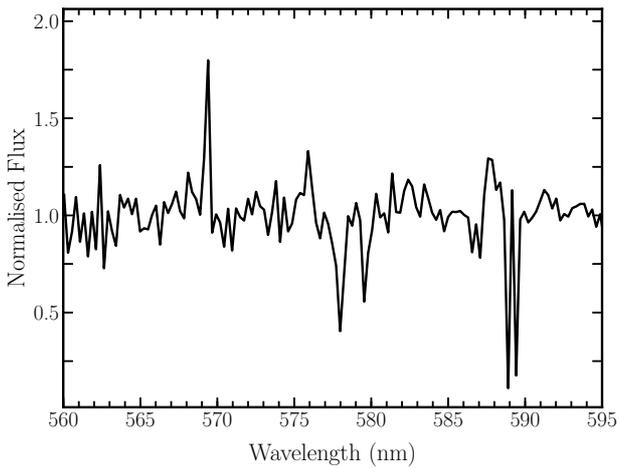}
 \caption{The spectrum of the northern companion to the central binary of Apep from 560 to 590\,nm.  C\,\textsc{iii}\,$\lambda$\,569.6\,nm emission  is detected in this wavelength range. The P\,Cygni He\,\textsc{i}\,$\lambda$\,587.6\,nm line is also detected. The spectral resolution of  this plot has been smoothed by a factor of 12 to increase the signal-to-noise. The strong Na\,\textsc{i} D interstellar doublet is also seen, as are prominent diffuse interstellar bands at 578 and 579.9\,nm.} 
\label{fig:north_com_vis_clines}
\end{center} 
\end{figure}

\begin{figure}
\begin{center}
\includegraphics[scale=0.45]{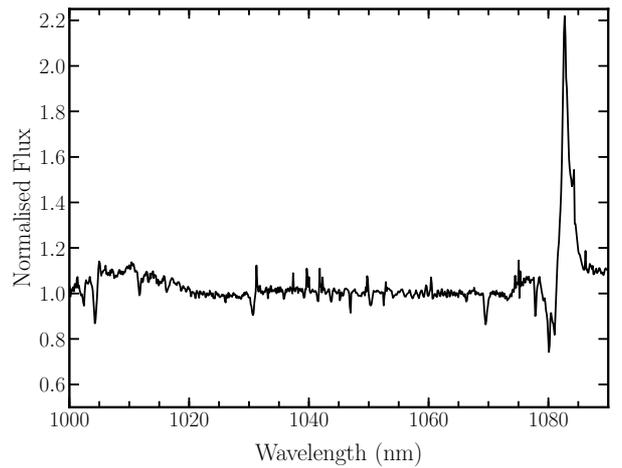}
 \caption{The spectrum of the northern companion to the central binary of Apep from 1000 to 1090\,nm, which displays the prominent He\,\textsc{i}\,$\lambda$\,1083\,nm line, plus weak He\,\textsc{ii}\,$\lambda$\,1012\,nm absorption and P$\delta$ at 1004.9\,nm.} 
\label{fig:north_com_nir}
\end{center}
\end{figure}

We provide the visual-red and near-infrared spectra of the northern companion to the central binary of Apep in Figure\,\ref{fig:north_com_vis}, \ref{fig:north_com_vis_clines}, and \ref{fig:north_com_nir}. The figures also highlight some of the diffuse interstellar bands and interstellar sodium and potassium lines, which are also present in the spectra of the central binary. 

Based on the $J$- to $K$-band spectra, \citet{2019NatAs...3...82C} suggested that the northern companion showed features similar to the B1\,Ia+ supergiant HD\,169454 \citep{2007A&A...465..993G}. With the detection of C\,\textsc{iii}\,$\lambda$\,569.6\,nm line emission, plus the separate components of the C\,\textsc{iii}\,$\lambda$\,971\,nm multiplet, the spectrum of the northern companion now conspicuously resembles that of late-O supergiants \citep{1999ApJ...511..374B}. In particular, the presence of weak He\,\textsc{ii}\,$\lambda$\,1012\,nm absorption and a P\,Cygni profile for the He\,\textsc{i}\,$\lambda$\,1083\,nm line confirm that the star is a late O supergiant, with close morphological similarities to HD\,151804 that is of O8\,Iaf spectral type \citep{1997A&A...317..532C, 1999ApJ...511..374B}. Additionally, the He\,\textsc{i}\,$\lambda$\,1083\,nm P\,Cygni line demonstrates the star has a terminal velocity of 1280\,$\pm$\,50\,km\,s$^{-1}$.

\section{Discussion}
\label{sec:discuss}

\subsection{Wind collision region}

The conditions and structure of the shock produced by the colliding winds in the central binary of Apep can be derived assuming the WC8+WC4-6 binary composition. There is evidence of two different pre-shock terminal wind velocities in the spectra. The first terminal wind velocity of $v_{\infty,\textrm{WN}} = 3500 \pm 100$\,km\,s$^{-1}$ is derived from fitting the P\,Cygni profile of the He\,\textsc{i}\,$\lambda$\,1083\,nm line \citep{1994MNRAS.269.1082E} shown in Figure\,\ref{fig:cent_binary_nir}. The second terminal wind velocity of $v_{\infty,\textrm{WC}} = 2100 \pm 200$\,km\,s$^{-1}$ is derived from the full-width at zero-intensity (FWZI) of the C\,\textsc{iii}\,$\lambda$\,569.6\,nm line presented in Figure\,\ref{fig:vis_clines}. The FWZI of the C\,\textsc{iii}\,$\lambda$\,569.6\,nm was measured by gridding the spectrum presented in Figure\,\ref{fig:vis_clines} to 0.15\,nm and measuring the width of the line 5\% above the average noise surrounding the line.

Since the C\,\textsc{iii}\,$\lambda$\,569.6\,nm line can only form in the WC8 star \citep{1968MNRAS.138..109S,1986ApJ...300..379T}, the faster wind measured from the P Cygni He\,\textsc{i}\,$\lambda$\,1083\,nm line is associated with the WN star in Apep. Such an association is also supported by the large FWHM of the He\,\textsc{ii}\,$\lambda$\,1012\,nm line. 

With the measured wind velocities of the two stars, it is possible to predict the momentum ratio for the colliding winds \citep{1992ApJ...389..635U}. Modelling the winds of both stars as spherical, the wind-momentum ratio $\eta$ is defined as

\begin{equation}
\eta = \dfrac{\dot{M}_{\textrm{WC}}v_{\infty,\textrm{WC}}}{\dot{M}_{\textrm{WN}}v_{\infty,\textrm{WN}}},
\end{equation} 

\noindent where $\dot{M}_{\textrm{WC}}$ and $\dot{M}_{\textrm{WN}}$ are the mass-loss rates of the WC8 and WN4-6b stars, respectively. Recent estimates of Galactic WC8 mass-loss rates span 10$^{-4.75}$ to 10$^{-4.35}$\,$M_{\odot}$\,yr$^{-1}$ \citep{2019A&A...621A..92S}, so we adopt a mass loss rate of 10$^{-4.5}$\,$M_{\odot}$\,yr$^{-1}$ for the WC8 component. For strong/broad lined WN4-6 stars, \citet{2019A&A...625A..57H} obtained a large mass-loss rate range of 10$^{-4.8}$ to 10$^{-3.8}$\,$M_{\odot}$\,yr$^{-1}$. Here we adopt a mass-loss rate of 10$^{-4.3}$\,$M_{\odot}$\,yr$^{-1}$ for the WN component. With these mass-loss rates and the assumption that the winds collide with the measured terminal velocities, the expected momentum ratio for the central binary of Apep is $\eta \sim 0.4$, with the shock bowed around the WC8 component. The value of $\eta \sim 0.4$ is in agreement with recent radio VLBI measurements \citep{marcote2020}. Such near-matched wind momenta is in stark contrast to standard WR+O colliding-wind binaries with $\eta \ll 0.1$, in which the WR wind overwhelms the wind of the O-type star. The momentum ratio of the central binary of Apep also implies that radiative breaking is unlikely to be at important in the shock region since the WN wind can not penetrate deeply into the acceleration region of the WC star's wind \citep{tuthill2008}.

The half-opening angle of shock $\theta$ (in radians) can be calculated using the formula provided by \citet{1993ApJ...402..271E}

\begin{equation}
\theta = 2.1\left(1 - \dfrac{\eta^{2/5}}{4}\right)\eta^{1/3}.
\end{equation}

\noindent For $\eta \sim 0.4$, the half-opening angle is expected to be $\approx$73\degree. This is somewhat larger than the half-opening angle of $\approx$60\degree~that is estimated from the spiral dust plume \citep{2019NatAs...3...82C}, consistent with the dust opening angle being smaller than the opening angle of the main shock \citep{tuthill2008} due to a switch in state of dust-production produced by an eccentric orbit \citep{han2020}. 

However, one significant problem with the CWB model thus far presented is that a momentum ratio of 0.4 predicts that the dust should inherit a shock-compressed velocity of $\approx$2500\,km\,s$^{-1}$ \citep[from Eqn. 29 of][]{Canto1996}, significantly higher than the observed dust expansion velocity of $\approx$600\,km\,s$^{-1}$ \citep{2019NatAs...3...82C}. 
The calculations performed above are predicated on the collision of two non-accelerating hypersonic spherical winds with the measured line-of-sight velocities, and that the postshock fluid is well mixed across the contact discontinuity separating the two radiative shocks. In contrast, Apep has been hypothesised to contain a dense, slow wind that emerges from the equator of one of the WR stars  \citep{2019NatAs...3...82C}. While the orbital period of the central binary is $\sim$100\,yr (Han et al. submitted), it is possible that the slow wind is present in the wind collision region, while the faster wind with one of the measured terminal velocities is escaping from the pole of that same WR star. 

To test if we can recover the measured dust expansion velocity of $\approx$600\,km\,s$^{-1}$ with the hypothesised slow and flattened equatorial wind being present in the wind-collision region, we assume the mass-loss rate is the same for both WR stars and $\eta = 0.4$. We find it is possible to recover a shock-compressed wind velocity along the dust shell consistent with the measured dust expansion velocity if the WC8 star produces a wind with a speed of $\approx$530\,km\,s$^{-1}$. Such a slow speed can be driven by near-critical rapid rotation of the WR star \citep{2006ApJ...637..914W,2014A&A...562A.118S}. For this calculation we modelled the WC8 star as having the hypothesised anisotropic mass-loss and slow, flattened wind since WC stars are known to dominate dust production in CWBs \citep[e.g.][]{2009MNRAS.395.2221W}.

Radio VLBI observations of Apep will be able to test if an half-opening angle of $\approx$73\degree~is suited to the central binary, and if the suggestion of the presence of the slow wind in the wind-collision region is accurate \citep{marcote2020}.

\subsection{Distance to Apep and association of northern companion}

With attribution of the spectral classes for all of the components of Apep, it is possible to estimate a distance to the source based on average luminosities of the different spectral classes. Apep has previously been suggested to be at a distance of $2.4^{+0.2}_{-0.5}$\,kpc, and definitely $<4.5$\,kpc based on kinematic information derived from diffuse interstellar bands \citep{2019NatAs...3...82C}. 

Since the northern companion closely resembles HD\,151804, and its spectrum likely has limited contamination from intrinsic dust, we estimate the extinction $E(B-V)$ by reddening the spectrum of the northern companion to roughly match HD\,151804. From that we estimate $E(B-V) \approx 4.2$, implying an absolute visual magnitude of $M_{\rm V} = -7.4$\,mag for a distance modulus of 11.9\,mag, using a standard extinction law with $R_{\rm V} = A_{\rm V}/E(B-V)$ = 3.1 and $V$-band magnitude of 17.8\,mag from \citet{2019NatAs...3...82C}. This compares closely with the absolute visual magnitude of $-7.2$ mag of HD~151804 (O8\,Iaf) from \citet{1997A&A...317..532C}.

For the central binary of Apep, we used models for WR\,137 (WC7+O) and WR\,110 (WN5-6b) to match the spectrum since the wind velocities of those stars are similar to those measured in the central binary. Again a reddening of $E(B-V) \approx 4.2$ generates a match to the combined spectra of the central binary, suggesting that the northern companion and central binary are associated. The combined absolute magnitude is $M_{\rm V}=-5.9$ mag for a distance modulus of 11.9\,mag, $R_{\rm V}$=3.1 and $V$-band magnitude of 19.0\,mag from \citep{2019NatAs...3...82C}. Under the reasonable assumption that each component contributes equally to the visual magnitude, as supported by the composite spectrum in Figure\,\ref{fig:comp}, each component would have $M_{\rm V}=-5.15$\,mag. Average absolute magnitudes for WC8 and WN4--6b stars in the WR visual band, $V_{\rm WR}$, are $-4.5\pm0.9$ mag and $-4.5\pm$0.6 mag, respectively. These values are consistent with a distance of $\sim$2.4 kpc derived by \citet{2019NatAs...3...82C}, and confirms the mismatch between the observed expansion velocity of the dusty spiral plume and spectroscopic wind in the system. In terms of the luminosity spread in both WR stars and O supergiants, a distance as close as 1.7\,kpc and as far as 2.4\,kpc are consistent with the photometry of Apep.

In summary, we favour a distance of 2.0$^{+0.4}_{-0.3}$\,kpc for Apep, and suggest the northern companion is likely associated with the central binary in a hierarchical triple system. Such an association is naturally favoured by the energetics of an O supergiant typical for its spectral class. However, we can not completely exclude the possibility that the northern companion is a rare chance line-of-sight alignment. If the northern companion is a chance alignment, such a configuration would help explain why there is no local heating of the dust near the star and the lack of a non-thermal source where the wind of the central binary impacts the northern companion's wind. The nearer distance of $\sim$1.7\,kpc is favoured photometrically if the northern companion is not associated with the central WR binary. Finally, we note that the \emph{Gaia} data release 2 \citep{2018A&A...616A...1G} parallax measurement for Apep is not accurate due to the multiplicity of the system \citep{2019NatAs...3...82C} and the influence of the complex extended dusty circumstellar environment. 

\section{Conclusion}
\label{sec:concl}

With the use of visual-red/near-infrared X-SHOOTER spectra, the three stellar components of the Apep system have been separated and decomposed into a central binary of two WR stars of subtypes WC8 and WN4-6b, together with an O8\,Iaf supergiant lying $\sim$0\arcsecfrac7 to the north. We contend that the central binary of Apep represents the strongest case of a classical WR+WR binary system in the Milky Way. The terminal wind velocity of the WC8 and WN4-6b stars are measured to be $2100 \pm 200$ and $3500 \pm 100$\,km\,s$^{-1}$, respectively.  If the mass-loss rates of the two WR stars are typical for their spectral class, the momentum ratio of the colliding wind is $\sim 0.4$. Such a large momentum ratio predicts a shock half-opening angle of $\approx$73\degree, $\approx$13\degree larger than that obtained by fitting models to the geometry of the spiral dust plume. 

However, this model does not fit with the observed proper motions of the dust plume. We can recover the momentum ratio and the measured slow dust expansion velocity if the mass-loss rate of both WR stars is assumed to be the identical but the wind from the WC8 star in the wind collision region has a velocity of $\approx$530\,km\,s$^{-1}$. The shock opening angle, and the presence of the slow wind is testable by way of radio VLBI observations \citep{marcote2020}. Finally, we show that the similar extinction measurements for the central binary and northern companion suggest a physical association, with the entire massive triple-system lying at 2.0$^{+0.4}_{-0.3}$\,kpc.

\section*{Acknowledgements}

Dedicated to P.~A.~C's father, Donald Crowther, who passed away due to COVID-19 during the preparation of this manuscript.

We thank the referee, Tomer Shenar (KU Leuven), for his insightful comments on the manuscript. J.~R.~C. thanks the Nederlandse Organisatie voor Wetenschappelijk Onderzoek (NWO) for support via the Talent Programme Veni grant. This study is based on observations collected at the European Organisation for Astronomical Research in the Southern Hemisphere under ESO programme 0103.D-0695(A).

This work was performed in part under contract with the Jet Propulsion Laboratory (JPL) funded by NASA through the Sagan Fellowship Program executed by the NASA Exoplanet Science Institute. B.~M. acknowledges support from the Spanish Ministerio de Econom\'ia y Competitividad (MINECO) under grant AYA2016-76012-C3-1-P.

This research made use of NASA's Astrophysics Data System, the \textsc{IPython} package \citep{PER-GRA:2007}; \textsc{SciPy} \citep{scipy};  \textsc{Matplotlib}, a \textsc{Python} library for publication quality graphics \citep{Hunter:2007}; \textsc{Astropy}, a community-developed core \textsc{Python} package for astronomy \citep{2013A&A...558A..33A}; and \textsc{NumPy} \citep{van2011numpy}.

%%%%%%%%%%%%%%%%%%%%%%%%%%%%%%%%%%%%%%%%%%%%%%%%%%



\bibliographystyle{mnras}
\bibliography{apep_xshooter.bbl} % if your bibtex file is called example.bib

\begin{thebibliography}{}
\makeatletter
\relax
\def\mn@urlcharsother{\let\do\@makeother \do\$\do\&\do\#\do\^\do\_\do\%\do\~}
\def\mn@doi{\begingroup\mn@urlcharsother \@ifnextchar [ {\mn@doi@}
  {\mn@doi@[]}}
\def\mn@doi@[#1]#2{\def\@tempa{#1}\ifx\@tempa\@empty \href
  {http://dx.doi.org/#2} {doi:#2}\else \href {http://dx.doi.org/#2} {#1}\fi
  \endgroup}
\def\mn@eprint#1#2{\mn@eprint@#1:#2::\@nil}
\def\mn@eprint@arXiv#1{\href {http://arxiv.org/abs/#1} {{\tt arXiv:#1}}}
\def\mn@eprint@dblp#1{\href {http://dblp.uni-trier.de/rec/bibtex/#1.xml}
  {dblp:#1}}
\def\mn@eprint@#1:#2:#3:#4\@nil{\def\@tempa {#1}\def\@tempb {#2}\def\@tempc
  {#3}\ifx \@tempc \@empty \let \@tempc \@tempb \let \@tempb \@tempa \fi \ifx
  \@tempb \@empty \def\@tempb {arXiv}\fi \@ifundefined
  {mn@eprint@\@tempb}{\@tempb:\@tempc}{\expandafter \expandafter \csname
  mn@eprint@\@tempb\endcsname \expandafter{\@tempc}}}

\bibitem[\protect\citeauthoryear{{Astropy Collaboration} et~al.,}{{Astropy
  Collaboration} et~al.}{2013}]{2013A&A...558A..33A}
{Astropy Collaboration} et~al., 2013, \mn@doi [\aap]
  {10.1051/0004-6361/201322068}, \href
  {http://adsabs.harvard.edu/abs/2013A%26A...558A..33A} {558, A33}

\bibitem[\protect\citeauthoryear{{Bartzakos}, {Moffat}  \&
  {Niemela}}{{Bartzakos} et~al.}{2001}]{bartzakos2001}
{Bartzakos} P.,  {Moffat} A.~F.~J.,   {Niemela} V.~S.,  2001, \mn@doi [\mnras]
  {10.1046/j.1365-8711.2001.04127.x}, \href
  {https://ui.adsabs.harvard.edu/abs/2001MNRAS.324...33B} {324, 33}

\bibitem[\protect\citeauthoryear{{Belczynski} et~al.,}{{Belczynski}
  et~al.}{2020}]{2017arXiv170607053B}
{Belczynski} K.,  et~al., 2020, \aap, \href
  {https://ui.adsabs.harvard.edu/abs/2017arXiv170607053B} {636, A104}

\bibitem[\protect\citeauthoryear{{Bohannan} \& {Crowther}}{{Bohannan} \&
  {Crowther}}{1999}]{1999ApJ...511..374B}
{Bohannan} B.,  {Crowther} P.~A.,  1999, \mn@doi [\apj] {10.1086/306647}, \href
  {http://adsabs.harvard.edu/abs/1999ApJ...511..374B} {511, 374}

\bibitem[\protect\citeauthoryear{{Callingham}, {Tuthill}, {Pope}, {Williams},
  {Crowther}, {Edwards}, {Norris}  \& {Kedziora-Chudczer}}{{Callingham}
  et~al.}{2019}]{2019NatAs...3...82C}
{Callingham} J.~R.,  {Tuthill} P.~G.,  {Pope} B.~J.~S.,  {Williams} P.~M.,
  {Crowther} P.~A.,  {Edwards} M.,  {Norris} B.,   {Kedziora-Chudczer} L.,
  2019, \mn@doi [Nature Astronomy] {10.1038/s41550-018-0617-7}, \href
  {https://ui.adsabs.harvard.edu/abs/2019NatAs...3...82C} {3, 82}

\bibitem[\protect\citeauthoryear{{Cant\'o}, {Raga}  \& {Wilkin}}{{Cant\'o}
  et~al.}{1996}]{Canto1996}
{Cant\'o} J.,  {Raga} A.~C.,   {Wilkin} F.~P.,  1996, \mn@doi [\apj]
  {10.1086/177820}, \href
  {https://ui.adsabs.harvard.edu/abs/1996ApJ...469..729C} {469, 729}

\bibitem[\protect\citeauthoryear{{Conti} \& {Massey}}{{Conti} \&
  {Massey}}{1989}]{1989ApJ...337..251C}
{Conti} P.~S.,  {Massey} P.,  1989, \mn@doi [\apj] {10.1086/167101}, \href
  {https://ui.adsabs.harvard.edu/abs/1989ApJ...337..251C} {337, 251}

\bibitem[\protect\citeauthoryear{{Crowther}}{{Crowther}}{2007}]{2007ARA&A..45..177C}
{Crowther} P.~A.,  2007, \mn@doi [\araa]
  {10.1146/annurev.astro.45.051806.110615}, \href
  {http://adsabs.harvard.edu/abs/2007ARA%26A..45..177C} {45, 177}

\bibitem[\protect\citeauthoryear{{Crowther} \& {Bohannan}}{{Crowther} \&
  {Bohannan}}{1997}]{1997A&A...317..532C}
{Crowther} P.~A.,  {Bohannan} B.,  1997, \aap, \href
  {https://ui.adsabs.harvard.edu/abs/1997A&A...317..532C} {317, 532}

\bibitem[\protect\citeauthoryear{{Crowther} \& {Smith}}{{Crowther} \&
  {Smith}}{1996}]{1996A&A...305..541C}
{Crowther} P.~A.,  {Smith} L.~J.,  1996, \aap, \href
  {http://adsabs.harvard.edu/abs/1996A%26A...305..541C} {305, 541}

\bibitem[\protect\citeauthoryear{{Crowther}, {De Marco}  \&
  {Barlow}}{{Crowther} et~al.}{1998}]{1998MNRAS.296..367C}
{Crowther} P.~A.,  {De Marco} O.,   {Barlow} M.~J.,  1998, \mn@doi [\mnras]
  {10.1046/j.1365-8711.1998.01360.x}, \href
  {https://ui.adsabs.harvard.edu/abs/1998MNRAS.296..367C} {296, 367}

\bibitem[\protect\citeauthoryear{{Dessart}, {Crowther}, {Hillier}, {Willis},
  {Morris}  \& {van der Hucht}}{{Dessart} et~al.}{2000}]{dessart2000}
{Dessart} L.,  {Crowther} P.~A.,  {Hillier} D.~J.,  {Willis} A.~J.,  {Morris}
  P.~W.,   {van der Hucht} K.~A.,  2000, \mn@doi [\mnras]
  {10.1046/j.1365-8711.2000.03399.x}, \href
  {https://ui.adsabs.harvard.edu/abs/2000MNRAS.315..407D} {315, 407}

\bibitem[\protect\citeauthoryear{{Detmers}, {Langer}, {Podsiadlowski}  \&
  {Izzard}}{{Detmers} et~al.}{2008}]{2008A&A...484..831D}
{Detmers} R.~G.,  {Langer} N.,  {Podsiadlowski} P.,   {Izzard} R.~G.,  2008,
  \mn@doi [\aap] {10.1051/0004-6361:200809371}, \href
  {http://adsabs.harvard.edu/abs/2008A%26A...484..831D} {484, 831}

\bibitem[\protect\citeauthoryear{{Eenens} \& {Williams}}{{Eenens} \&
  {Williams}}{1994}]{1994MNRAS.269.1082E}
{Eenens} P.~R.~J.,  {Williams} P.~M.,  1994, \mn@doi [\mnras]
  {10.1093/mnras/269.4.1082}, \href
  {http://adsabs.harvard.edu/abs/1994MNRAS.269.1082E} {269, 1082}

\bibitem[\protect\citeauthoryear{{Eichler} \& {Usov}}{{Eichler} \&
  {Usov}}{1993}]{1993ApJ...402..271E}
{Eichler} D.,  {Usov} V.,  1993, \mn@doi [\apj] {10.1086/172130}, \href
  {https://ui.adsabs.harvard.edu/abs/1993ApJ...402..271E} {402, 271}

\bibitem[\protect\citeauthoryear{{Gaia Collaboration} et~al.,}{{Gaia
  Collaboration} et~al.}{2018}]{2018A&A...616A...1G}
{Gaia Collaboration} et~al., 2018, \mn@doi [\aap]
  {10.1051/0004-6361/201833051}, \href
  {https://ui.adsabs.harvard.edu/abs/2018A%26A...616A...1G} {616, A1}

\bibitem[\protect\citeauthoryear{{Groh}, {Damineli}  \& {Jablonski}}{{Groh}
  et~al.}{2007}]{2007A&A...465..993G}
{Groh} J.~H.,  {Damineli} A.,   {Jablonski} F.,  2007, \mn@doi [\aap]
  {10.1051/0004-6361:20066401}, \href
  {http://adsabs.harvard.edu/abs/2007A%26A...465..993G} {465, 993}

\bibitem[\protect\citeauthoryear{{Hamann}, {Koesterke}  \&
  {Wessolowski}}{{Hamann} et~al.}{1995}]{hamann1995}
{Hamann} W.~R.,  {Koesterke} L.,   {Wessolowski} U.,  1995, \aaps, \href
  {https://ui.adsabs.harvard.edu/abs/1995A&AS..113..459H} {113, 459}

\bibitem[\protect\citeauthoryear{{Hamann} et~al.,}{{Hamann}
  et~al.}{2019}]{2019A&A...625A..57H}
{Hamann} W.~R.,  et~al., 2019, \mn@doi [\aap] {10.1051/0004-6361/201834850},
  \href {https://ui.adsabs.harvard.edu/abs/2019A&A...625A..57H} {625, A57}

\bibitem[\protect\citeauthoryear{{Han}, {Tuthill}, {Soulain}, {Callingham},
  {Williams}, {Crowther}, {Pope}  \& {Marcote}}{{Han} et~al.}{subm}]{han2020}
{Han} Y.,  {Tuthill} P.,  {Soulain} A.,  {Callingham} J.~R.,  {Williams} P.,
  {Crowther} P.,  {Pope} B.,   {Marcote} B.,  subm., \mnras

\bibitem[\protect\citeauthoryear{{Hill}, {Moffat}, {St-Louis}  \&
  {Bartzakos}}{{Hill} et~al.}{2000}]{hill2000}
{Hill} G.~M.,  {Moffat} A.~F.~J.,  {St-Louis} N.,   {Bartzakos} P.,  2000,
  \mn@doi [\mnras] {10.1046/j.1365-8711.2000.03705.x}, \href
  {https://ui.adsabs.harvard.edu/abs/2000MNRAS.318..402H} {318, 402}

\bibitem[\protect\citeauthoryear{{Hillier}}{{Hillier}}{1989}]{1989ApJ...347..392H}
{Hillier} D.~J.,  1989, \mn@doi [\apj] {10.1086/168127}, \href
  {https://ui.adsabs.harvard.edu/abs/1989ApJ...347..392H} {347, 392}

\bibitem[\protect\citeauthoryear{{Howarth} \& {Schmutz}}{{Howarth} \&
  {Schmutz}}{1992}]{Howarth1992}
{Howarth} I.~D.,  {Schmutz} W.,  1992, \aap, \href
  {https://ui.adsabs.harvard.edu/abs/1992A&A...261..503H} {261, 503}

\bibitem[\protect\citeauthoryear{Hunter}{Hunter}{2007}]{Hunter:2007}
Hunter J.~D.,  2007, Computing In Science \& Engineering, 9, 90

\bibitem[\protect\citeauthoryear{Jones, Oliphant, Peterson  \& Others}{Jones
  et~al.}{2001}]{scipy}
Jones E.,  Oliphant T.,  Peterson P.,   Others 2001, {SciPy}: Open source
  scientific tools for Python, \url {http://www.scipy.org/}

\bibitem[\protect\citeauthoryear{{Koenigsberger}, {Morrell}, {Hillier},
  {Gamen}, {Schneider}, {Gonz{\'a}lez-Jim{\'e}nez}, {Langer}  \&
  {Barb{\'a}}}{{Koenigsberger} et~al.}{2014}]{Koenigsberger2014}
{Koenigsberger} G.,  {Morrell} N.,  {Hillier} D.~J.,  {Gamen} R.,  {Schneider}
  F. R.~N.,  {Gonz{\'a}lez-Jim{\'e}nez} N.,  {Langer} N.,   {Barb{\'a}} R.,
  2014, \mn@doi [\aj] {10.1088/0004-6256/148/4/62}, \href
  {https://ui.adsabs.harvard.edu/abs/2014AJ....148...62K} {148, 62}

\bibitem[\protect\citeauthoryear{{Lamers}, {Maeder}, {Schmutz}  \&
  {Cassinelli}}{{Lamers} et~al.}{1991}]{1991ApJ...368..538L}
{Lamers} H.~J.~G.~L.~M.,  {Maeder} A.,  {Schmutz} W.,   {Cassinelli} J.~P.,
  1991, \mn@doi [\apj] {10.1086/169717}, \href
  {https://ui.adsabs.harvard.edu/abs/1991ApJ...368..538L} {368, 538}

\bibitem[\protect\citeauthoryear{{Langer}}{{Langer}}{1992}]{langer1992}
{Langer} N.,  1992, \aap, \href
  {https://ui.adsabs.harvard.edu/abs/1992A&A...265L..17L} {265, L17}

\bibitem[\protect\citeauthoryear{{Levan}, {Crowther}, {de Grijs}, {Langer},
  {Xu}  \& {Yoon}}{{Levan} et~al.}{2016}]{2016SSRv..202...33L}
{Levan} A.,  {Crowther} P.,  {de Grijs} R.,  {Langer} N.,  {Xu} D.,   {Yoon}
  S.-C.,  2016, \mn@doi [\ssr] {10.1007/s11214-016-0312-x}, \href
  {http://adsabs.harvard.edu/abs/2016SSRv..202...33L} {202, 33}

\bibitem[\protect\citeauthoryear{{Luehrs}}{{Luehrs}}{1997}]{luehrs1997}
{Luehrs} S.,  1997, \mn@doi [\pasp] {10.1086/133907}, \href
  {https://ui.adsabs.harvard.edu/abs/1997PASP..109..504L} {109, 504}

\bibitem[\protect\citeauthoryear{{Maeder}}{{Maeder}}{1987}]{maeder1987}
{Maeder} A.,  1987, \aap, \href
  {https://ui.adsabs.harvard.edu/abs/1987A&A...178..159M} {178, 159}

\bibitem[\protect\citeauthoryear{{Mandel} \& {de Mink}}{{Mandel} \& {de
  Mink}}{2016}]{mandel2016}
{Mandel} I.,  {de Mink} S.~E.,  2016, \mn@doi [\mnras] {10.1093/mnras/stw379},
  \href {https://ui.adsabs.harvard.edu/abs/2016MNRAS.458.2634M} {458, 2634}

\bibitem[\protect\citeauthoryear{{Marchant}, {Langer}, {Podsiadlowski},
  {Tauris}  \& {Moriya}}{{Marchant} et~al.}{2016}]{marchant2016}
{Marchant} P.,  {Langer} N.,  {Podsiadlowski} P.,  {Tauris} T.~M.,   {Moriya}
  T.~J.,  2016, \mn@doi [\aap] {10.1051/0004-6361/201628133}, \href
  {https://ui.adsabs.harvard.edu/abs/2016A&A...588A..50M} {588, A50}

\bibitem[\protect\citeauthoryear{{Marchenko} et~al.,}{{Marchenko}
  et~al.}{2003}]{Marchenko2003}
{Marchenko} S.~V.,  et~al., 2003, \mn@doi [\apj] {10.1086/378154}, \href
  {http://adsabs.harvard.edu/abs/2003ApJ...596.1295M} {596, 1295}

\bibitem[\protect\citeauthoryear{{Marcote}, {Callingham}, {De Becker},
  {Edwards}, {Han}, {Schulz}, {Stevens}  \& {Tuthill}}{{Marcote}
  et~al.}{subm}]{marcote2020}
{Marcote} B.,  {Callingham} J.~R.,  {De Becker} M.,  {Edwards} P.,  {Han} Y.,
  {Schulz} R.,  {Stevens} J.,   {Tuthill} P.,  subm., \mnras

\bibitem[\protect\citeauthoryear{{Martins}, {Depagne}, {Russeil}  \&
  {Mahy}}{{Martins} et~al.}{2013}]{martins2013}
{Martins} F.,  {Depagne} E.,  {Russeil} D.,   {Mahy} L.,  2013, \mn@doi [\aap]
  {10.1051/0004-6361/201321282}, \href
  {https://ui.adsabs.harvard.edu/abs/2013A&A...554A..23M} {554, A23}

\bibitem[\protect\citeauthoryear{{Massey} \& {Grove}}{{Massey} \&
  {Grove}}{1989}]{1989ApJ...344..870M}
{Massey} P.,  {Grove} K.,  1989, \mn@doi [\apj] {10.1086/167854}, \href
  {http://adsabs.harvard.edu/abs/1989ApJ...344..870M} {344, 870}

\bibitem[\protect\citeauthoryear{{Modigliani} et~al.,}{{Modigliani}
  et~al.}{2010}]{2010SPIE.7737E..28M}
{Modigliani} A.,  et~al., 2010, {The X-shooter pipeline}.
p. 773728, \mn@doi{10.1117/12.857211}

\bibitem[\protect\citeauthoryear{{Monnier}, {Tuthill}  \& {Danchi}}{{Monnier}
  et~al.}{1999}]{1999ApJ...525L..97M}
{Monnier} J.~D.,  {Tuthill} P.~G.,   {Danchi} W.~C.,  1999, \mn@doi [\apjl]
  {10.1086/312352}, \href {http://adsabs.harvard.edu/abs/1999ApJ...525L..97M}
  {525, L97}

\bibitem[\protect\citeauthoryear{P\'erez \& Granger}{P\'erez \&
  Granger}{2007}]{PER-GRA:2007}
P\'erez F.,  Granger B.~E.,  2007, \mn@doi [Computing in Science and
  Engineering] {10.1109/MCSE.2007.53}, 9, 21

\bibitem[\protect\citeauthoryear{{Pittard}}{{Pittard}}{2009}]{pittard2009}
{Pittard} J.~M.,  2009, \mn@doi [\mnras] {10.1111/j.1365-2966.2009.14857.x},
  \href {https://ui.adsabs.harvard.edu/abs/2009MNRAS.396.1743P} {396, 1743}

\bibitem[\protect\citeauthoryear{{Rate} \& {Crowther}}{{Rate} \&
  {Crowther}}{2020}]{2020MNRAS.493.1512R}
{Rate} G.,  {Crowther} P.~A.,  2020, \mn@doi [\mnras] {10.1093/mnras/stz3614},
  \href {https://ui.adsabs.harvard.edu/abs/2020MNRAS.493.1512R} {493, 1512}

\bibitem[\protect\citeauthoryear{Rosslowe \& Crowther}{Rosslowe \&
  Crowther}{2018}]{Rosslowe2017}
Rosslowe C.~K.,  Crowther P.~A.,  2018, \mn@doi [Monthly Notices of the Royal
  Astronomical Society] {10.1093/mnras/stx2103}, 473, 2853

\bibitem[\protect\citeauthoryear{{Sander}, {Hamann}, {Todt}, {Hainich},
  {Shenar}, {Ramachandran}  \& {Oskinova}}{{Sander}
  et~al.}{2019}]{2019A&A...621A..92S}
{Sander} A.~A.~C.,  {Hamann} W.~R.,  {Todt} H.,  {Hainich} R.,  {Shenar} T.,
  {Ramachandran} V.,   {Oskinova} L.~M.,  2019, \mn@doi [\aap]
  {10.1051/0004-6361/201833712}, \href
  {https://ui.adsabs.harvard.edu/abs/2019A&A...621A..92S} {621, A92}

\bibitem[\protect\citeauthoryear{{Schnurr}, {Casoli}, {Chen{\'e}}, {Moffat}  \&
  {St-Louis}}{{Schnurr} et~al.}{2008}]{2008MNRAS.389L..38S}
{Schnurr} O.,  {Casoli} J.,  {Chen{\'e}} A.~N.,  {Moffat} A.~F.~J.,
  {St-Louis} N.,  2008, \mn@doi [\mnras] {10.1111/j.1745-3933.2008.00517.x},
  \href {https://ui.adsabs.harvard.edu/abs/2008MNRAS.389L..38S} {389, L38}

\bibitem[\protect\citeauthoryear{{Shenar}, {Hamann}  \& {Todt}}{{Shenar}
  et~al.}{2014}]{2014A&A...562A.118S}
{Shenar} T.,  {Hamann} W.-R.,   {Todt} H.,  2014, \mn@doi [\aap]
  {10.1051/0004-6361/201322496}, \href
  {http://adsabs.harvard.edu/abs/2014A%26A...562A.118S} {562, A118}

\bibitem[\protect\citeauthoryear{{Shenar} et~al.,}{{Shenar}
  et~al.}{2019}]{shenar2019}
{Shenar} T.,  et~al., 2019, \mn@doi [\aap] {10.1051/0004-6361/201935684}, \href
  {https://ui.adsabs.harvard.edu/abs/2019A&A...627A.151S} {627, A151}

\bibitem[\protect\citeauthoryear{{Smith}}{{Smith}}{1968}]{1968MNRAS.138..109S}
{Smith} L.~F.,  1968, \mn@doi [\mnras] {10.1093/mnras/138.1.109}, \href
  {http://adsabs.harvard.edu/abs/1968MNRAS.138..109S} {138, 109}

\bibitem[\protect\citeauthoryear{{Smith}, {Shara}  \& {Moffat}}{{Smith}
  et~al.}{1990}]{1990ApJ...358..229S}
{Smith} L.~F.,  {Shara} M.~M.,   {Moffat} A. F.~J.,  1990, \mn@doi [\apj]
  {10.1086/168978}, \href
  {https://ui.adsabs.harvard.edu/abs/1990ApJ...358..229S} {358, 229}

\bibitem[\protect\citeauthoryear{{Smith}, {Shara}  \& {Moffat}}{{Smith}
  et~al.}{1996}]{1996MNRAS.281..163S}
{Smith} L.~F.,  {Shara} M.~M.,   {Moffat} A. F.~J.,  1996, \mn@doi [\mnras]
  {10.1093/mnras/281.1.163}, \href
  {https://ui.adsabs.harvard.edu/abs/1996MNRAS.281..163S} {281, 163}

\bibitem[\protect\citeauthoryear{{Song}, {Meynet}, {Maeder}, {Ekstr{\"o}m}  \&
  {Eggenberger}}{{Song} et~al.}{2016}]{song2016}
{Song} H.~F.,  {Meynet} G.,  {Maeder} A.,  {Ekstr{\"o}m} S.,   {Eggenberger}
  P.,  2016, \mn@doi [\aap] {10.1051/0004-6361/201526074}, \href
  {https://ui.adsabs.harvard.edu/abs/2016A&A...585A.120S} {585, A120}

\bibitem[\protect\citeauthoryear{{Torres}, {Conti}  \& {Massey}}{{Torres}
  et~al.}{1986}]{1986ApJ...300..379T}
{Torres} A.~V.,  {Conti} P.~S.,   {Massey} P.,  1986, \mn@doi [\apj]
  {10.1086/163811}, \href
  {https://ui.adsabs.harvard.edu/abs/1986ApJ...300..379T} {300, 379}

\bibitem[\protect\citeauthoryear{{Tuthill}, {Monnier}  \& {Danchi}}{{Tuthill}
  et~al.}{1999}]{1999Natur.398..487T}
{Tuthill} P.~G.,  {Monnier} J.~D.,   {Danchi} W.~C.,  1999, \mn@doi [\nat]
  {10.1038/19033}, \href {http://adsabs.harvard.edu/abs/1999Natur.398..487T}
  {398, 487}

\bibitem[\protect\citeauthoryear{{Tuthill}, {Monnier}, {Lawrance}, {Danchi},
  {Owocki}  \& {Gayley}}{{Tuthill} et~al.}{2008}]{tuthill2008}
{Tuthill} P.~G.,  {Monnier} J.~D.,  {Lawrance} N.,  {Danchi} W.~C.,  {Owocki}
  S.~P.,   {Gayley} K.~G.,  2008, \mn@doi [\apj] {10.1086/527286}, \href
  {http://adsabs.harvard.edu/abs/2008ApJ...675..698T} {675, 698}

\bibitem[\protect\citeauthoryear{{Usov}}{{Usov}}{1992}]{1992ApJ...389..635U}
{Usov} V.~V.,  1992, \mn@doi [\apj] {10.1086/171236}, \href
  {https://ui.adsabs.harvard.edu/abs/1992ApJ...389..635U} {389, 635}

\bibitem[\protect\citeauthoryear{Van Der~Walt, Colbert  \& Varoquaux}{Van
  Der~Walt et~al.}{2011}]{van2011numpy}
Van Der~Walt S.,  Colbert S.~C.,   Varoquaux G.,  2011, Computing in Science \&
  Engineering, 13, 22

\bibitem[\protect\citeauthoryear{{Vernet} et~al.,}{{Vernet}
  et~al.}{2011}]{2011A&A...536A.105V}
{Vernet} J.,  et~al., 2011, \mn@doi [\aap] {10.1051/0004-6361/201117752}, \href
  {https://ui.adsabs.harvard.edu/abs/2011A&A...536A.105V} {536, A105}

\bibitem[\protect\citeauthoryear{{Williams}, {van der Hucht}, {Pollock},
  {Florkowski}, {van der Woerd}  \& {Wamsteker}}{{Williams}
  et~al.}{1990}]{1990MNRAS.243..662W}
{Williams} P.~M.,  {van der Hucht} K.~A.,  {Pollock} A.~M.~T.,  {Florkowski}
  D.~R.,  {van der Woerd} H.,   {Wamsteker} W.~M.,  1990, \mnras, \href
  {http://adsabs.harvard.edu/abs/1990MNRAS.243..662W} {243, 662}

\bibitem[\protect\citeauthoryear{{Williams} et~al.,}{{Williams}
  et~al.}{2009a}]{williams2009}
{Williams} P.~M.,  et~al., 2009a, \mn@doi [\mnras]
  {10.1111/j.1365-2966.2009.14664.x}, \href
  {https://ui.adsabs.harvard.edu/abs/2009MNRAS.395.1749W} {395, 1749}

\bibitem[\protect\citeauthoryear{{Williams}, {Rauw}  \& {van der
  Hucht}}{{Williams} et~al.}{2009b}]{2009MNRAS.395.2221W}
{Williams} P.~M.,  {Rauw} G.,   {van der Hucht} K.~A.,  2009b, \mn@doi [\mnras]
  {10.1111/j.1365-2966.2009.14681.x}, \href
  {https://ui.adsabs.harvard.edu/abs/2009MNRAS.395.2221W} {395, 2221}

\bibitem[\protect\citeauthoryear{{Willis}}{{Willis}}{1982}]{1982MNRAS.198..897W}
{Willis} A.~J.,  1982, \mn@doi [\mnras] {10.1093/mnras/198.4.897}, \href
  {https://ui.adsabs.harvard.edu/abs/1982MNRAS.198..897W} {198, 897}

\bibitem[\protect\citeauthoryear{{Woosley} \& {Heger}}{{Woosley} \&
  {Heger}}{2006}]{2006ApJ...637..914W}
{Woosley} S.~E.,  {Heger} A.,  2006, \mn@doi [\apj] {10.1086/498500}, \href
  {http://adsabs.harvard.edu/abs/2006ApJ...637..914W} {637, 914}

\bibitem[\protect\citeauthoryear{{Zhekov}, {Tomov}, {Gawronski}, {Georgiev},
  {Borissova}, {Kurtev}, {Gagn{\'e}}  \& {Hajduk}}{{Zhekov}
  et~al.}{2014}]{2014MNRAS.445.1663Z}
{Zhekov} S.~A.,  {Tomov} T.,  {Gawronski} M.~P.,  {Georgiev} L.~N.,
  {Borissova} J.,  {Kurtev} R.,  {Gagn{\'e}} M.,   {Hajduk} M.,  2014, \mn@doi
  [\mnras] {10.1093/mnras/stu1880}, \href
  {https://ui.adsabs.harvard.edu/abs/2014MNRAS.445.1663Z} {445, 1663}

\bibitem[\protect\citeauthoryear{{de Mink}, {Cantiello}, {Langer}, {Pols},
  {Brott}  \& {Yoon}}{{de Mink} et~al.}{2009}]{demink2009}
{de Mink} S.~E.,  {Cantiello} M.,  {Langer} N.,  {Pols} O.~R.,  {Brott} I.,
  {Yoon} S.~C.,  2009, \mn@doi [\aap] {10.1051/0004-6361/200811439}, \href
  {https://ui.adsabs.harvard.edu/abs/2009A&A...497..243D} {497, 243}

\bibitem[\protect\citeauthoryear{{de Mink}, {Langer}, {Izzard}, {Sana}  \& {de
  Koter}}{{de Mink} et~al.}{2013}]{2013ApJ...764..166D}
{de Mink} S.~E.,  {Langer} N.,  {Izzard} R.~G.,  {Sana} H.,   {de Koter} A.,
  2013, \mn@doi [\apj] {10.1088/0004-637X/764/2/166}, \href
  {https://ui.adsabs.harvard.edu/abs/2013ApJ...764..166D} {764, 166}

\bibitem[\protect\citeauthoryear{{van Hoof}}{{van
  Hoof}}{2018}]{2018Galax...6...63V}
{van Hoof} P. A.~M.,  2018, \mn@doi [Galaxies] {10.3390/galaxies6020063}, \href
  {https://ui.adsabs.harvard.edu/abs/2018Galax...6...63V} {6, 63}

\makeatother
\end{thebibliography}

% Don't change these lines
\bsp	% typesetting comment
\label{lastpage}
\end{document}